\documentclass[final]{raa}            

\usepackage{graphicx,times}             
\usepackage{natbib}
\usepackage{amssymb,amsmath}
\bibpunct{(}{)}{;}{a}{}{,}

\newcommand{\cm}{{~\rm cm}}

\newcommand{\km}{{~\rm km}}
\newcommand{\s}{{~\rm s}}

\newcommand{\g}{{~\rm g}}

\newcommand{\erg}{{~\rm erg}}
\newcommand{\yr}{{~\rm yr}}

\newcommand{\pc}{{~\rm pc}}

\newcommand{\AU}{{~\rm AU}}

\newcommand{\days}{{~\rm days}}

\begin{document}

   \title{A common envelope jets supernova (CEJSN) impostor scenario for fast blue optical transients
}

   \volnopage{Vol.0 (20xx) No.0, 000--000}      
   \setcounter{page}{1}          

   \author{Noam Soker
      \inst{1}
   }

   \institute{Department of Physics, Technion, Haifa, 3200003, Israel; {\it soker@physics.technion.ac.il}\\
\vs\no
   {\small Received~~20xx month day; accepted~~20xx~~month day}}

\abstract{
I propose a new scenario, the \textit{polar common envelope jets supernova (CEJSN) impostor} scenario, to account for AT2018cow-like fast blue optical transients (FBOTs). 
The polar CEJSN impostor scenario evolves through four main phases. (1) A red supergiant (RSG) star expands to tidally interact with a neutron star (NS) companion (or a black hole). The interaction increases the RSG mass loss rate to form a circumstellar matter (CSM) halo to $r \simeq 0.1 \pc$. (2) Shortly before the onset of a common envelope evolution (CEE) and about a year before explosion the NS accretes mass from the RSG envelope and launches jets that inflate two opposite lobes in the CSM within $\approx 100 \AU$. (3) The NS-RSG system enters a CEE phase during which the system ejects most of the envelope mass in a dense equatorial outflow. (4) At the termination of the CEE the leftover envelope forms a circumbinary disk around the NS-core system. The NS accretes mass from the circumbinary disk and launches energetic jets that when collide with the fronts of the CSM lobes power an FBOT event. The low mass of the jets-lobes interaction zones and their large distance, of about $100 \AU$, from the center account for the fast transient. In the future the core collapses to form a second NS. In the far future the two NS might merge. I suggest that FBOTs and similar fast transients are CEJSN impostors which compose a large fraction of the progenitors of NS-NS merger binaries.
 }

 \authorrunning{N. Soker}            
   \titlerunning{A CEJSN impostor scenario for FBOTs}  
   
      \maketitle
\section{Introduction} 
\label{sec:intro}

In the common envelope jet supernova (CEJSN) scenario the source of explosion energy is the accretion of gas from a red supergiant (RSG) star onto a neutron star (NS) or a black hole (BH), hereafter NS/BH, that spirals-in inside the RSG, first inside the envelope and then inside the RSG core (e.g., \citealt{Papishetal2015, SokerGilkis2018, Gilkisetal2019, GrichenerSoker2019a, LopezCamaraetal2019, LopezCamaraetal2020MN, Soker2021}). The accretion process proceeds via an accretion disk that launches energetic jets that explode the star.  

In case the NS/BH spirals-in inside the envelope of the giant star but does not enter (does not merge with) the core, the event is \textit{a CEJSN impostor} \citep{Gilkisetal2019}. 

The jets that the NS/BH launches at late times interact with the remaining RSG envelope and with the circumstellar matter (CSM) that the earlier common envelope evolution (CEE) ejected. The interaction excites two shock waves, the forward shock that expands into the ambient gas (the envelope and/or the CSM) and the reverse shock that shocks the outflowing jets' gas. The cocoon is the hot post-shock regions of the two shocks, which have a contact discontinuity between them. The hot cocoon radiates some of its energy (e.g., \citealt{Schreieretal2021}), leading to a bright transient event that mimic a core collapse supernova (CCSN). I do not always explicitly mention the cocoon. However, it should be taken for granted that any jet interaction with the envelope and/or with the CSM leads to the formation of a cocoon (e.g., \citealt{LopezCamaraetal2019, Schreieretal2021}). For further discussion of the accretion rate in the CEJSN scenario see \cite{Gricheneretal2021} and \cite{Hilleletal2022}. 

Because they mimic CCSNe and have a rich variety of properties, { e.g., the companion  can be a NS or a BH with a range of masses, the RSG envelope mass and radius can vary by factors of about an order of magnitude, the core mass and composition can vary, and the eccentricity of the orbit before the strong interaction can be from zero to a high value } (e.g., \citealt{SokeretalGG2019}), CEJSN and CEJSN impostor events are very likely to be behind several types of transient events that CCSNe cannot account for. \cite{Thoneetal2011} explain the unusual gamma-ray burst GRB~101225A by a NS that merged with a helium star. The enigmatic supernovae iPTF14hls \citep{Arcavietal2017} and SN~2020faa \citep{Yangetal2020} might be CEJSN events \citep{SokerGilkis2018}. 
Adopting the CEJSN scenario and processes from these earlier studies, \cite{Dongetal2021} proposed that the luminous radio transient VT~J121001+495647 was a CEJSN event. CEJSNe might account also for some processes that require extreme conditions, like some fraction of the r-process nucleosynthesis \citep{GrichenerSoker2019a, GrichenerSoker2019b, Gricheneretal2022} and, for a CEJSN impostor with a BH companion, the formation of a fraction of the very high-energy neutrinos \citep{GrichenerSoker2021}. 

In this study I concentrate on AT2018cow-like fast blue optical transients (FBOTs). 
{AT2018cow-like FBOTs are fast-rising, only few days (e.g., \citealt{Prenticeeta2018, Perleyetal2019}), bright transients, i.e, might be brighter than superluminous CCSNe (e.g., \citealt{Marguttietal2019}). They display high velocities of $\ga 0.1 c$ with a total kinetic energy of $\simeq 10^{51}-10^{52} \erg$ (e.g., \citealt{Coppejansetal2020}), 
with hydrogen lines (e.g., \citealt{Marguttietal2019}).  They tend to occur in star-forming galaxies (e.g., \citealt{Prenticeeta2018}). They might display rapid X-ray variability (e.g., \citealt{Pashametal2021, Yaoetal2022}) and have a dense CSM (e.g., \citealt{NayanaChandra2021, Brightetal2021}). In section \ref{sec:ObservatinalProperties} I return to discuss in more details the properties of FBOTs, including their uncertain rate. } 

{ In the CEJSN scenario a NS/BH that was formed in an earlier CCSN of the initially more massive star in a binary system enters the envelope of a RSG that evolved from the initially less massive star of the binary. The main powering of the event comes from the very energetic jets that the NS/BH that spirals-in inside the envelope and then inside the core of the RSG star launches. The very powerful jets is the main difference from the many types of binary interactions with and without CEE (e.g., \citealt{Hanetal2020}), from cases where a main sequence companion launches jets in a CEE (e.g., \citealt{Shiberetal2016}), and from the more closely related NS/BH scenarios of CEE without jets} (e.g., \citealt{FryerWoosley1998, ZhangFryer2001, BarkovKomissarov2011, Thoneetal2011, Chevalier2012, Schroderetal2020}). 

In the polar CEJSN channel of the CEJSN scenario that \cite{SokeretalGG2019} constructed, the early jets remove most of the envelope gas along the polar directions. Late jets then expand almost freely to large distances such that they might account for the fast rise and decline of FBOTs, as well as for the very high outflow velocities of up to $> 0.1c$ (e.g., \citealt{Marguttietal2019, Perleyetal2019, Coppejansetal2020}). Jet-CSM interaction must take place to convert kinetic energy to thermal energy of the post-shock jet and CSM gas (the cocoon). This hot gas radiates a large fraction of its energy to power a bright event. 

The enigmatic AT~2018cow event \citep{Prenticeeta2018} and similar FBOTs have promoted the development of several scenarios for AT~2018cow-like FBOTs (e.g., \citealt{Liuetal2018, FoxSmith2019, Kuinetal2019, LyutikovToonen2019, Marguttietal2019, Quataertetal2019, Yuetal2019, Leungetal2020, Mohanetal2020, PiroLu2020, UnoMaeda2020, Kremeretal2021, Xiangetal2021, ChenShen2022, Gottliebetal2022}). In the present study I do not compare the different scenarios with each other but rather present the \textit{polar-CEJSN impostor} channel (section \ref{sec:PolarImpostor}) and its new ingredients (section \ref{sec:NewIngredients}). While \cite{SokeretalGG2019} presented a polar CEJSN scenario, I here present a polar CEJSN impostor scenario, i.e., the NS/BH does not enter (nor destroy) the RSG core. The scenario I propose here does not replace the one \cite{SokeretalGG2019} proposed, but rather adds another channel to the CEJSN and CEJSN impostor scenarios.  
In section \ref{sec:ObservatinalProperties} I discuss how the polar CEJSN impostor scenario might account for the observational properties of the AT2018cow-like FBOTs. { I summarize in section \ref{sec:Summary}. } 

\section{The polar-CEJSN impostor scenario} 
\label{sec:PolarImpostor}

I here describe the general polar CEJSN impostor scenario. I will present the details of some ingredients in section \ref{sec:NewIngredients}.
I emphasise that the main difference between the polar CEJSN impostor that I propose here for FBOTs and the polar CEJSN scenario for FBOTs that \cite{SokeretalGG2019} proposed is that, as the `impostor' in the name implies, in the present scenario the NS does not enter (does not merge with) the core of the RSG star. 

{ The system will experience the CEJSN impostor scenario if the system manages to remove the entire RSG envelope before the NS reaches the core. This in turns depends on the radius and mass of the RSG and on the eccentricity of the orbit when the strong binary interaction of the NS and the RSG starts. A lower envelope mass, a higher RSG radius, and a higher eccentricity favour envelope removal hence CEJSN impostor evolution. However, there is yet no quantitative study of these effects. }

The main evolutionary phases after the formation of a binary system of a NS and a RSG star are as follows. 
\begin{enumerate}
\item \textit{Pre-CEE.} The NS perturbs the RSG envelope by tidal forces, that both deform the envelope and spin it up. As a result of that the mass loss rate increases by possibly { up to about an order of magnitude. } This wind forms the CSM halo (section \ref{subsec:CSM}). 
\item \textit{The onset of the CEE.} As the NS enters the envelope (the onset of the CEE) it starts to accrete mass via an accretion disk and launches jets { (see \citealt{LopezCamaraetal2020MN} for simulations and \citealt{Hilleletal2022} for discussion of the accretion process). } During this phase the jets interact with the dense wind and shape two opposite lobes along the bipolar directions, as observed in some planetary nebulae (section \ref{subsec:CSM}). Since the NS is just about to enter the envelope and accretion is similar to a Roche Lobe overflow, the jets manage to expand and accelerate gas along the two opposite polar directions.  
This takes place about a year before the explosion, which is the duration of the CEE phase $\tau_{\rm CEE}$ { (about the dynamical time on the surface of the RSG star, e.g., \citealt{Lauetal2022}), } and might lead to a faint precursor at  $t_{\rm PreC} \simeq - \tau_{\rm CEE} \approx - 1 \yr$ relative to the FBOT event itself.  
\item \textit{The main CEE phase.} During the main CEE phase the NS spirals-in inside the RSG envelope, { and when it removes the entire envelope it ends at a final orbital radius from the RSG core that I mark as $a_{\rm NC}$. } The jets do not penetrate the envelope and most mass is ejected near the equatorial plane and up to mid-latitudes (e.g., \citealt{Schreieretal2021}). The average jets' power is $\approx 10^{43} \erg \s^{-1}$. The outflow does not destroy the polar lobes, but rather mainly adds mass to the dense equatorial outflow.
\item \textit{Accretion from a circumbinary disk as the energy source of the explosion.} Following \cite{KashiSoker2011}, I assume that at the end of the main CEE phase most of the leftover envelope mass forms a circumbinary disk around the NS-core binary system (section \ref{subsec:CBD}). I further assume that there are two opposite openings in the envelope along the polar directions (funnels; e.g., \citealt{Soker1992, Zouetal2020}). Therefore, at the end of the CEE the jets that the NS launches as it accretes hydrogen-rich gas from the circumbinary disk mange to escape the star. The `polar' in the name of the polar CEJSN scenario that \cite{SokeretalGG2019} proposed and in the present polar CEJSN impostor scenario comes from the (almost) empty polar directions inside the RSG envelope that allow the jets to propagate to large distances. The difference is that in the polar CEJSN the NS launches jets as it accretes mass from the destroyed massive core of the RSG, leading to (1) a very energetic event with a kinetic energy of $\ga 10^{52} \erg$, much above the kinetic energy of typical CCSNe ($\approx {\rm few} \times 10^{50} - {\rm few} \times 10^{51} \erg$), and (2) hydrogen-poor jets. The typical explosion energy of the FBOTs that I consider here is as of typical CCSNe and the fast outflow is hydrogen rich (section \ref{sec:ObservatinalProperties}), which bring me to propose that the NS does not enter the RSG core. Hence the `impostor' in the name of the scenario.     
\item \textit{Late accretion phase from the circumbinary disk.}
{ The accretion process depletes the circumbinary disk, which is the reservoir for the accreted mass, and therefore the accretion rate decreases. } The accretion rate decreases over a timescale of weeks, namely, it might continue for weeks (section \ref{subsec:CBD}). This explains the late X-ray rapid variability (section \ref{sec:ObservatinalProperties}).
\item \textit{Far-future events.} This scenario predicts that in the future (up to about hundreds of thousands of years in the future) the RSG core experience a striped-CCSN (type Ib or type Ic CCSN) event, leaving behind a second NS. At an even later time the two NSs might merge as in the channel-I CEE scenario that \cite{VignaGomezetal2018} study for NS-NS merger events. 
\end{enumerate}

\section{New ingredients} 
\label{sec:NewIngredients}

I here describe the new ingredients of the polar-CEJSN impostor scenario with respect to the polar-CEJSN scenario that \cite{SokeretalGG2019} introduced. 
The main difference, as the `impostor' in the name implies, is that the NS does not enter (does not merge with) the core of the RSG. Note, however, that the bipolar CSM structure that I describe next might be also the CSM structure in the polar CEJSN scenario. 

Several emission sources and properties as well as the dense equatorial CSM of the polar-CEJSN impostor scenario are similar to the general picture that \cite{Marguttietal2019} describe. { These similar properties include a bipolar structure of the interaction, a central engine, late X-ray variability that results from changes in the power of the central engine, a fast polar outflow that explains the absorption lines, and an equatorial outflow that accounts for the $\simeq {\rm several} \times 1000 \km \s^{-1}$ broad hydrogen emission lines at later times. Another process that \cite{SokeretalGG2019} adopted from \cite{Marguttietal2019} is the receding of the photosphere from the fast polar ejecta to the slower equatorial ejecta, which might explain the transition from X-ray to UV/visible/IR
emission. } Here I further note that some processes and properties that I describe here are also similar to some processes in the FBOT scenario that \cite{Gottliebetal2022} develop. There are, however, some basic differences between the scenarios, in particular that in the polar-CEJSN impostor scenario the NS is an old one, rather than a newly born NS or BH { as in the scenario of \cite{Gottliebetal2022}, } and that in the polar CEJSN impostor scenarios the CSM contains two opposite lobes along the polar directions. The strong interaction of the jets that yields the early  optical emission is with the CSM, rather than with the stellar envelope that \cite{Gottliebetal2022} consider. At late times, i.e., after a few weeks, some emission properties, like radio emission, are similar as in both scenarios the interaction of the jets is with the CSM.    
   
\subsection{Pre-explosion bipolar CSM} 
\label{subsec:CSM}
 
\begin{figure}
\begin{center}
\includegraphics[trim=20.5cm 13.0cm 29.0cm 2.50cm,scale=0.86]{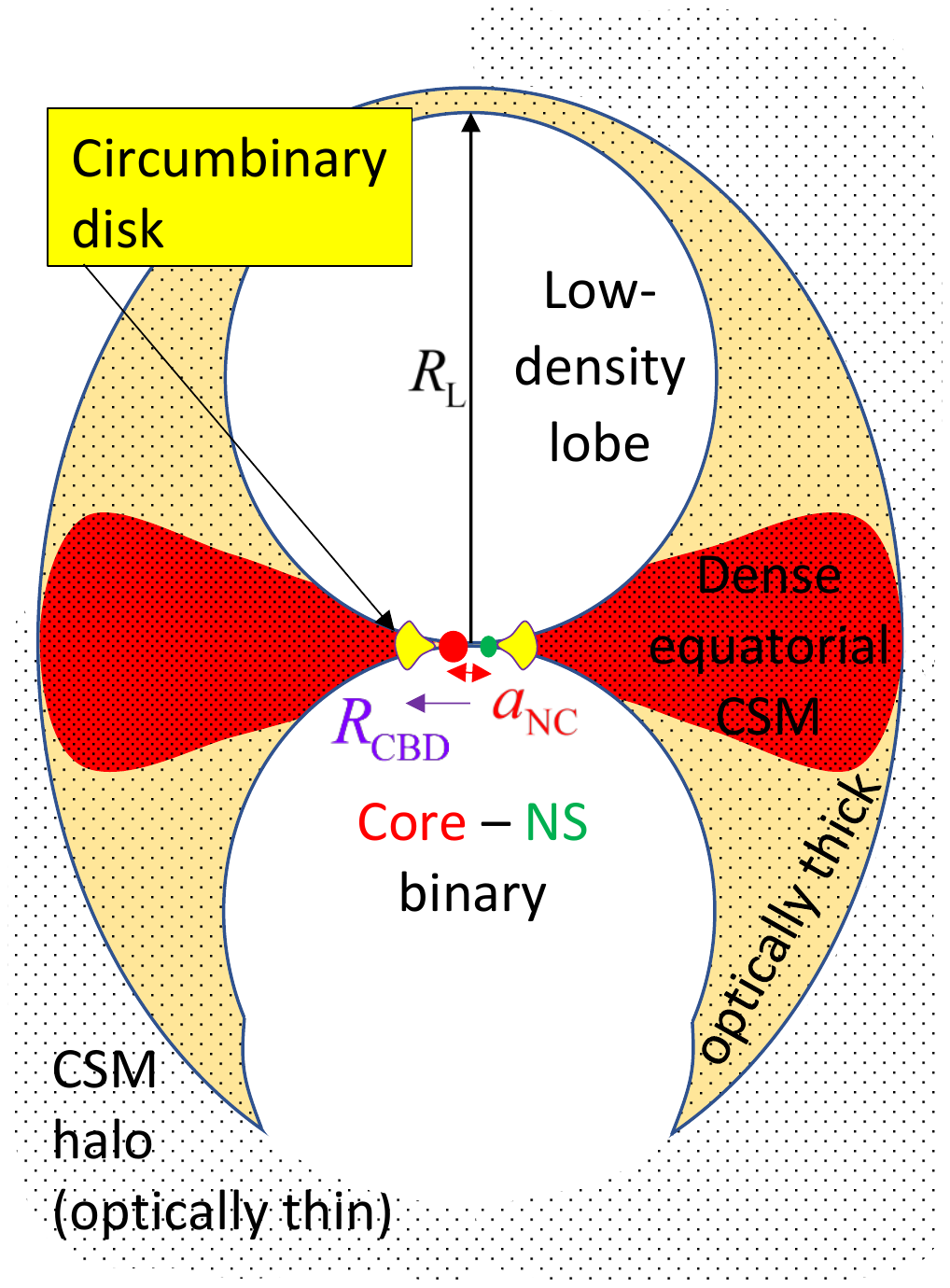} 
\caption{A schematic drawing (not to scale) of the CSM just before the explosion, { which is also very shortly after the system exits the CEE. } The plane of the figure is the meridional plane that momentarily contains the RSG core (red filled-circle) and the NS (green dot). There is an axial-symmetry about a vertical axis in the figure through the center of mass of the core-NS binary system (the symmetry axis coincides with the $R_{\rm L}$ arrow in the figure). There is also a mirror symmetry about the equatorial plane. However, in this figure the upper lobe represents the case of the two lobes being closed by the optically-thick shell, while the lower lobe represents the case where the two lobes are open. { The main structural components are the low-density lobes (in white), the optically thick shell that encloses the two lobes from all or most sides (dotted-yellow), the dense equatorial CSM (red), the circumbinary disk (the two yellow zones in the inner part), and the CSM halo that extends beyond the boundary of the figure (dotted-white area). 
$R_{\rm L} \approx 10^{15} \cm$ is the length of one lobe along the polar direction, $a_{\rm NC} \simeq 2 R_\odot$ is the post-CEE orbital separation of the NS and the core, and $R_{\rm CBD} \simeq {\rm several} \times a_{\rm NC}$ is the radius of the post-CEE circumbinary disk. }
}
\label{fig:FBOTsCSM}
\end{center}
\end{figure}

I take the pre-explosion CSM structure just after the NS reaches close to the RSG core to be similar to that of some bipolar planetary nebulae, e.g., the Owl Nebula (NGC 3587; e.g., \citealt{GarciaDiazetal2018}). The structure of this type of planetary nebulae contains two opposite low-density lobes (the white ellipse in the upper half of Fig. \ref{fig:FBOTsCSM}) inside a spherical or an elliptical shell (black dots with yellow background in Fig. \ref{fig:FBOTsCSM}). { Jets inflate such lobes. In the present scenario the NS launches jets just before it enters the envelope of the RSG star. At the same time its interaction with the envelope increases the mass loss rate in the RSG wind by about three orders of magnitude to be $\approx 0.01 M_\odot \yr^{-1}$. These jets that are active for several months inflate the lobes as they interact with the intensive wind from the RSG star. } The shell is optically thick. 
 
The inflation of these two lobes compresses the equatorial gas (e.g., \citealt{AkashiSoker2008}), adding to the action of the gravity of the companion, here a NS, and jets at the onset of the CEE (e.g. \citealt{Shiberetal2019}). I schematically draw this CSM structure in Fig. \ref{fig:FBOTsCSM}. Note that there is a mirror symmetry about the equatorial plane, but in Fig. \ref{fig:FBOTsCSM} the upper lobe represents the case where the two lobes are closed by the optically-thick shell (the walls of the lobes) while the lower lobe represents the case where the two lobes are open. { The ratio of the energy of the pre-CEE jets that inflate the bubbles to the density of the dense wind is the main factor that determines the shape of the lobes (other parameters include the opening angle of the jets and the wind velocity). Very energetic jets will break out from the dense wind and form open lobes.  }
    
The duration of the CEE phase, $\tau_{\rm CEE}$, from the onset of the CEE until the compact companion spirals-in deeply into the giant envelope is about the orbital (Keplerian) time on the surface of the giant (e.g., \citealt{GlanzPerets2021}). In the present study the orbital time
on the surface of a RSG is about a year. During that time the CSM lobes that the jets expel reach a distance of 
\begin{equation}
R_{\rm L } \simeq 10^{15} 
\left( \frac{\tau_{\rm CEE}}{1 \yr} \right)
\left( \frac{v_{\rm L}}{300 \km \s^{-1}} \right) \cm ,
\label{eq:RL}    
\end{equation}
where for the expansion velocity of the lobe, $v_{\rm L}$, I take a velocity larger by about a factor of three than the escape velocity from the RSG star,
$v_{\rm L} \simeq 3 v_{\rm esc}(R_{\rm RSG})$, where $R_{\rm RSG} \simeq {\rm few} \times AU$ is the radius of the RSG. The equatorial outflow is slower, and closer to the escape velocity from the RSG.

I emphasise the following properties.
\newline
(1) The formation of the two opposite polar lobes of the CSM can take place before the CEE. Examples are the Red Rectangle nebula around a post-asymptotic giant branch binary system and the Homunculus nebula around the massive binary system Eta Carinae. Both binary systems have a bipolar CSM around them, e.g., \cite{Cohenetal1975} and \cite{Smith2006}, respectively, but did not enter a full CEE. I suggest the same in the case of some FBOTs.

(2) Likely, there is a more spherical CSM and of lower density at much larger distances, as observed in many planetary nebulae (e.g. \citealt{Corradietal2003}), e.g., the Owl Nebula. This is the CSM halo (black dots with a white background in Fig. \ref{fig:FBOTsCSM}), which is optically thin. 

(3) It is during the CEE that the binary system ejects the massive and dense equatorial CSM. 

(4) During most of the CEE the jets do not break out from the envelope (e.g., \citealt{GrichenerSoker2021,Hilleletal2022}). They can break out only at the beginning of the CEE {just before the NS enters the very dense parts of the envelope, or at the end of the CEE when the CEE cleared the polar directions (e.g., \citealt{Soker2019TerminateCEE}). }

(5) Because the companion star spins-up the envelope during the CEE, in some cases the fast envelope rotation leads to the opening of a funnel along the two polar directions (e.g., \citealt{Soker1992, Zouetal2020}).  This ensures that the jets expand almost freely before they encounter the dense parts of the lobes. The two opposite openings (funnels) in the polar directions give this scenario the name `polar CEJSN' \citep{SokeretalGG2019} or polar CEJSN impostor (present study). 
The funnels in the polar directions are another property that distinguishes the polar CEJSN scenario from the scenario that \cite{Gottliebetal2022} propose in which the jets strongly interact with the envelope. 

(6) It is possible that the fronts of the two lobes are open. I schematically present this possibility in the lower lobe in Fig. \ref{fig:FBOTsCSM}.

\subsection{Post CEE accretion from a circumbinary disk} 
\label{subsec:CBD}

\cite{Marguttietal2019} find that the central power source (“engine”) of AT~2018cow should release a total energy of $\simeq 10^{50}-10^{51.5} \erg$ over a characteristic timescale of $\approx 10^3-10^5 \s$. They also deduce that the ejecta mass is $M_{\rm ej} \simeq 0.1-1 M_\odot$ and that it contains hydrogen and helium, but a limited $^{56}$Ni mass of $M_{\rm Ni} <0.04 M_\odot$. The ejecta velocity span a large range, from $v_{\rm ej} \la 0.01 \km \s^{-1}$ to $v_{\rm ej} \simeq 0.2c$. I here raise the possibility that the NS launches these outflows as it accretes mass from a circumbinary disk at the end of the CEE inside the RSG envelope, i.e., the RSG core is still intact. 

\cite{KashiSoker2011} argue that $\eta_{\rm CBD} \simeq 0.01-0.1$ of the common envelope might remain bound in a circumbinary disk after the compact companion ends the CEE inside the giant envelope (later it might enter the core or destroy the core). A dynamically stable circumbinary disk extends from $R_{\rm CBD} \simeq 2.5 a_{\rm NC}$ to $R_{\rm CBD} \simeq {\rm several} \times a_{\rm NC}$, where $a_{\rm NC}$ is the orbital separation of the NS and the core (assuming a circular orbit). Here I do not require the circumbinary disk to be dynamically stable, and its inner boundary will be much closer to the NS-core binary system because the NS-core systems has just emerged from the CEE. 
I assume that the NS launches fast jets at velocity $v_{\rm j}$ shortly after the CEE phase and that the jets carry a fraction of $\eta_{\rm 2j} \simeq 0.05-0.1$ of the circumbinary mass. For an RSG envelope mass $M_{\rm env}$ the mass in the two jets and their energy are then 
\begin{equation}
M_{\rm 2j} = 0.01 
\left( \frac{\eta_{2j}}{0.05} \right)
\left( \frac{\eta_{\rm CBD}}{0.02} \right)
\left( \frac{M_{\rm env}}{10 M_\odot} \right) M_\odot, 
\label{eq:JetsMass}
\end{equation}
\begin{equation}
E_{\rm 2j} = 10^{51} 
\left( \frac{M_{\rm 2j}}{0.01 M_\odot} \right)
\left( \frac{v_{\rm j}}{10^5 \km \s^{-1}} \right)
\erg,
\label{eq:JetsEnergy}
\end{equation}
respectively. This energy is about equal to the jets' energy in the scenario of \cite{Gottliebetal2022}, but here the jets are mainly baryonic while \cite{Gottliebetal2022} consider relativistic jets. 

The duration of the accretion process is about the viscosity time of the circumbinary disk, which is $\approx 10-100$ times the Keplerian orbital time of the disk around the binary. The accretion time period of the entire post-CEE circumbinary disk, from  its inner boundary to its outer boundary, is  
\begin{eqnarray}
\tau_{\rm CBD} & \approx \left( 1- 100 \right)    
\left( \frac{a_{\rm NC}}{2 R_\odot} \right)^{3/2}  \nonumber
\\ & \times
\left( \frac{M_{\rm core} +M_{\rm NS}}{7 M_\odot} \right)^{-1/2} \days
\label{eq:CSMtime}
\end{eqnarray}
where $M_{\rm core}$ is the mass of the RSG core and I took the inner and outer boundaries of the circumbinary disk to be $R_{\rm CBD,in} \simeq a_{\rm NC}$ and $R_{\rm CBD,out} \simeq 5 a_{\rm NC}$, respectively. I expect a high accretion rate at the first few days, { $\dot M_{\rm acc} \approx 0.01-0.1 M_\odot {\rm day}^{-1}$, } that slowly decreases with timescales of weeks to months to { $\dot M_{\rm acc} \approx 10^{-4}-10^{-3} M_\odot {\rm day}^{-1}$, } but not necessarily monotonically. { The decline in the mass accretion rate might at best launch weak jets that cannot propagate to large distances. This might be related to the finding of \cite{Bietenholzet2020B} that there are no observed long-lived relativistic jets in in AT2018cow.}

The phase of mass accretion from the circumbinary disk is a short phase that precedes the BB mass transfer phase in the channel-I CEE scenario that \cite{VignaGomezetal2018} describe (their figure 5) for the formation of double NSs that later merge by gravitational wave emission. Namely, I here suggest that some of the AT1018cow-like FBOTs are progenitors of binary NS systems that much later merge by gravitational waves. In the channel-I scenario that \cite{VignaGomezetal2018} describe the NS ends at $a_{\rm NC} \simeq 1-{\rm few} \times R_\odot$ from the helium core. The helium core masses in most cases they consider for channel-I are $M_{\rm core} \simeq 4-7 M_\odot$. They consider a phase of stable mass transfer from the core to the NS, which implies $a_{\rm NC} \ga 1.5 R_\odot$ (e.g., \citealt{Taurisetal2015}).  Since I do not require here a stable mass transfer at a phase after the explosion, I allow for smaller orbital separations of even $a_{\rm NC} \la 1 R_\odot$. 

\subsection{The interaction of the jets with the lobes} 
\label{subsec:Interaction}

\begin{figure}
\begin{center}
\includegraphics[trim=20.5cm 13.0cm 29.0cm 2.50cm,scale=0.86]{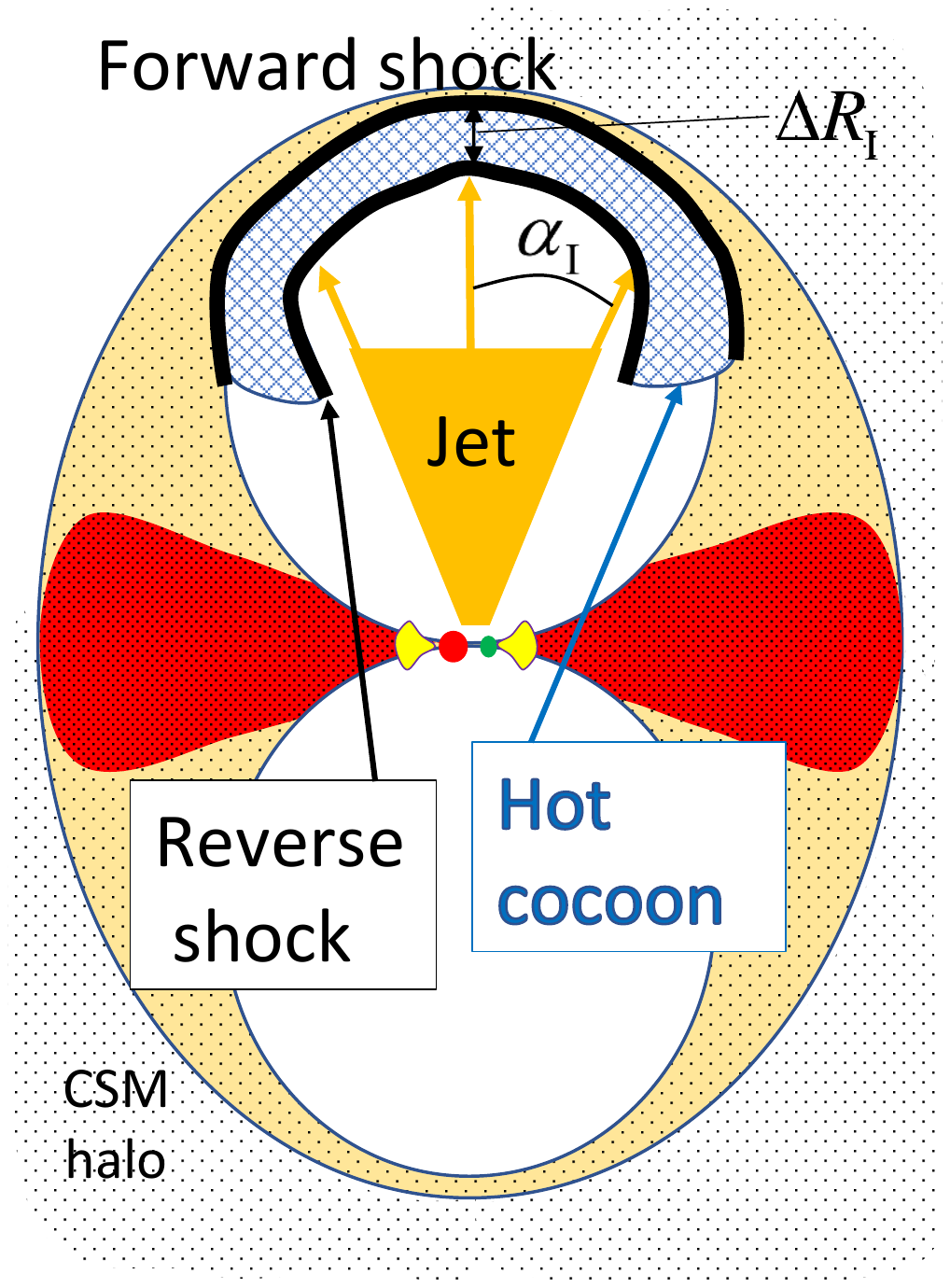} 
\caption{A schematic drawing (not to scale) of the interaction of the post-CEE hydrogen-rich jets with the lobes before shock break-out. The NS launches the jets as it orbits the RSG core and accretes from the circumbinary disk.
Interaction takes place in both lobes, but the drawing is only in the upper one. The elliptical CSM shell (yellow background) is optically thick, while the CSM halo is optically thin. }
\label{fig:FBOTsInteractionfig2}
\end{center}
\end{figure}

As the jets expand they cool adiabatically. To channel a large fraction of the jets' kinetic energy to radiation they must collide with an ambient gas. In the polar CEJSN impostor scenario  the jets collide with the bipolar CSM as I schematically present in Fig. \ref{fig:FBOTsInteractionfig2} (only for one lobe). Due to the orbital motion of the NS and the interaction of the jets with some tenuous gas in the lobes' interior (the lobes are not completely empty) I expect the jets not to be collimated, and they might even be wide. { Specifically, interaction of jets with a dense close (to the launching point) gas can collimate the jets. Here the polar directions are almost empty, and the tenuous gas along the polar directions will be accelerate and entrained by the jets to a wider polar outflow (wide jets). In addition, the orbital motion prevents the build-up of a dense gas very close to the NS. }  As well, when they collide with the dense front of the lobe the jets might be slower and more massive than at their origin near the NS, { as they entrain the tenuous gas along the polar directions. }  

\begin{figure}
\begin{center}
\includegraphics[trim=20.5cm 12.2cm 29.0cm 4.0cm,scale=0.86]{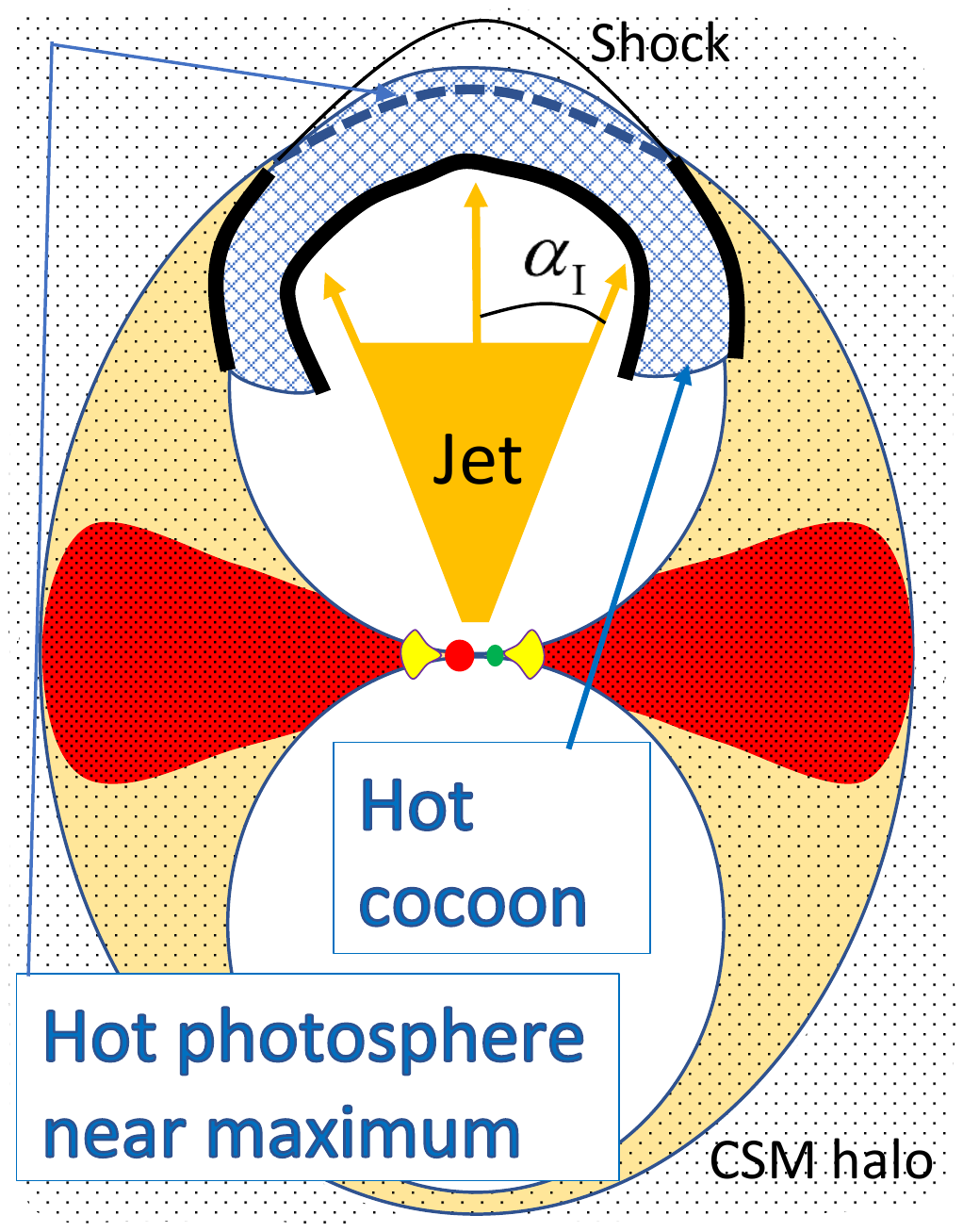} 
\caption{Similar to Fig. \ref{fig:FBOTsInteractionfig2} but after shock break-out and before maximum optical luminosity. The dashed blue line marks the photosphere on one side (another exists on the opposite lobe but is not drawn). The hatched blue volume outside the photosphere that once was a hot cocoon is already cold due to radiative cooling. }
\label{fig:FBOTsInteractionfig3}
\end{center}
\end{figure}

Consider then that a jet collides with the front of the lobe over an angle $\alpha_{\rm I}$ (Fig. \ref{fig:FBOTsInteractionfig2}) such that the solid-angle of the two interaction regions (one in each lobe) is $\Omega_{\rm I} =4 \pi (1- \cos \alpha_{\rm I})$. The jets shock the front of the close lobe, and when the forward shock breaks out from the lobe there is a hot cocoon of width $\Delta R_{\rm I}$.
At that time the photosphere starts to move inward with respect to the mass of the cocoon, and a weaker shock continues to propagate into the optically-thin halo, as I schematically show in Fig. \ref{fig:FBOTsInteractionfig3}.

The combined volume of the two hot cocoons that break out from the lobe in the jet-lobe interaction regions is $V_{\rm I} \simeq \Omega_{\rm I} R^2_{\rm L} \Delta R_{\rm I}$. The mass inside each of the two hot cocoons includes the mass of the jet, the mass that the jet drags with it as it expands through the lobe's interior, and the mass of the dense shell of the lobe within the interaction region. This mass will be several times the original mass of the jet. I scale this mass with $M_{\rm I} \simeq 0.1 M_\odot$.
The photon diffusion time from this region is  
\begin{eqnarray}
\begin{aligned} 
t_{\rm diff} & \simeq \frac{3 \tau \Delta R_{\rm I}}{c} 
\simeq  2   
\left( \frac{M_{\rm I}}{0.1 M_\odot} \right)
\left( \frac{\kappa}{0.1 \cm^2 \g^{-1}} \right) 
\\ & \times 
\left( \frac{R_{\rm L}}{10^{15} \cm} \right)^{-1}
\left( \frac{\Delta R_{\rm I}}{0.3 R_{\rm L}} \right) 
\left( \frac{\Omega_{\rm I}}{\pi}\right)^{-1}
\days,
\end{aligned}
\label{eq:tdiff}
\end{eqnarray}
where $\tau=\rho_{\rm I} \kappa \Delta R_{\rm I}$ is the optical depth of the interaction region, $\rho_{\rm I}$ is the density of the cocoon, and $\kappa$ is the opacity. Equation (\ref{eq:tdiff}) gives the timescale for the variation of the luminosity and photosphere size at rise and early decline.  

\begin{figure}
\begin{center}
\includegraphics[trim=20.5cm 12.2cm 29.0cm 4.0cm,scale=0.86]{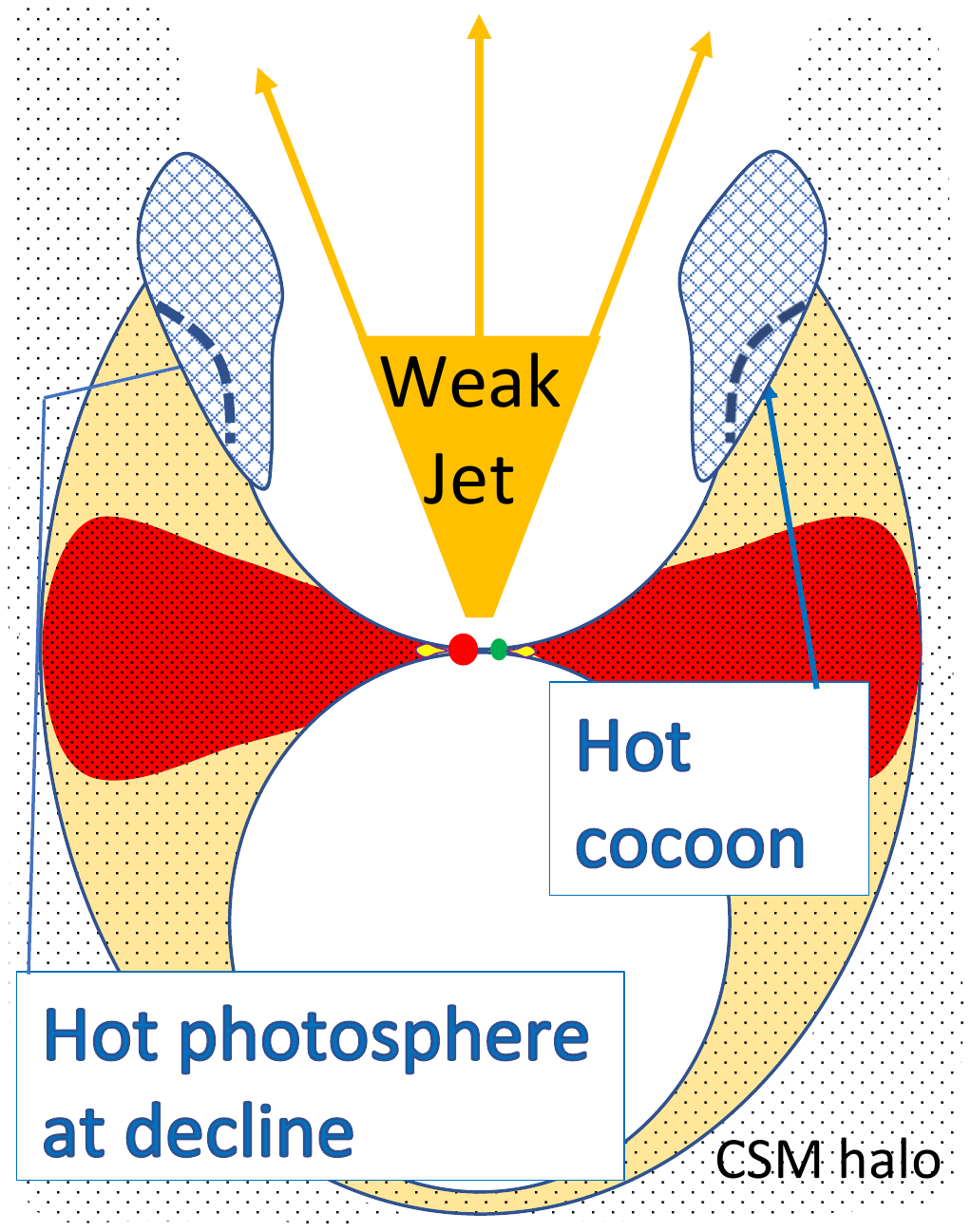} 
\caption{Similar to Fig. \ref{fig:FBOTsInteractionfig3} but during the decline phase of the light curve as the photosphere recedes into lower latitudes of the walls of the lobes. The jets are weaker now and have already cleaned the polar directions. Note that now an observer at a high enough latitude can see the other side of the lobe, as well as the central region.
(As before, the lower lobe shows the structure before the interaction with jets for comparison, although interaction takes place also there). }
\label{fig:FBOTsInteractionfig4}
\end{center}
\end{figure}

Due to this geometry a spherically-symmetric model of the photosphere will yield a smaller distance from the center, depending on the viewing angle of the observer. Approximately, the inferred radius of a spherical model be $R_{\rm ph,sph} \simeq R_{\rm L} (\Omega_{\rm I}/4 \pi)^{1/2}$.  For example, the earliest photosphere radius that \cite{Perleyetal2019} deduce for AT2018cow is $R_{\rm ph,sph} = 8 \times 10^{14} \cm$. For the scaling I use here of $\Omega_{\rm I} = \pi$ the lobe radius is $ R_{\rm L} \simeq 1.6 \times 10^{15} \cm$. \cite{Perleyetal2019} find that the radius of the photosphere that they calculate decreases from the first time they calculate this radius. I attribute this monotonic decrease of the photosphere radius to the structure of the lobes, as I draw schematically in Fig. \ref{fig:FBOTsInteractionfig4}. At the decline phase the low density regions near the symmetry axis are already optically thin and the jets clean these regions. The shocks that run into the walls of the lobes at lower latitudes form hot cocoons there that continue to radiate, but from a smaller and smaller photosphere areas.  

With the simple scenario that I describe here I cannot follow to late times when the luminosity and photosphere area have dropped by a large factor. { The reason is that the photosphere area becomes small, and therefore it is sensitive to the initial conditions and the jet-lobe interaction that I cannot follow with analytical means at late times. } \cite{Perleyetal2019} find that at $t \simeq 40 \days$ the luminosity and photosphere area of AT2018cow have dropped by about three and two orders of magnitude, respectively. At these late times the optical emission properties depend on small-scale interaction of the jets with left-over RSG envelope gas near the polar directions. As well, there might be a slow disk-wind from the circumbinary disk. 

At late time an observer that is not at too low latitudes can see the other side of the lobes' walls, and then the central region, namely, the core-NS binary system and the circumbinary disk (that has its mass  decreasing due to accretion and disk-wind). 

{ I end this section by mentioning the case of open lobes. In this case the jets interact with less mass of the lobes, but still interact with a mass. I expect that in that case the fast outflow (jets) will be more pronounced and that the channelling of the kinetic energy of the jets to radiation will be less efficient. Therefore, the FBOT will be fainter. The exact properties depend strongly on the density of the gas inside the lobes and the opening angle of the lobes, as well as the jets' properties of course.  }
 
\section{Accounting for observational properties} 
\label{sec:ObservatinalProperties}

In this section I discuss some of the properties of AT~2018-like FBOTs alongside the processes in the polar CEJSN impostor scenario that might account for these properties. 

In comparing the theoretical expectations with observations one should note the following. Firstly, there are several processes that require quantitative studies to determine more accurate values and to check the suggestions that I raise here. In the present study I propose the scenario but there are hydrodynamical simulations, radiative transfer calculations, and population synthesis studies to conduct in the future.

Secondly, the parameter space of the polar CEJSN impostor scenario is large.  
\cite{Coppejansetal2020} points out the diversity of FBOTs. Indeed, the polar CEJSN impostor scenario has properties that can change from one FBOT to another, including the CSM structure (compare the two lobes in Fig. \ref{fig:FBOTsCSM}), the mass in the circumbinary disk, the companion type (a NS or a BH), and the evolutionary phase of the RSG and its mass at the onset of the CEE. 

In what follows I consider the compact object that spirals-in inside the RSG star to be a NS. However, in some cases it might be a BH. 

\subsection{Star-forming galaxies} 
\label{subsec:Star-forming galaxies}
AT2018cow-like FBOTs tend to occur in star-forming galaxies, namely, come from massive stars (e.g., \citealt{Prenticeeta2018, Perleyetal2019, Lymanetal2020, Perleyetal2021}). \cite{SokeretalGG2019} noted that from their population synthesis study \cite{Mapellietal2018} find that in the local Universe NS-NS mergers tend to occur shortly after star formation. Since many CEJSN events lead to NS-NS close binary systems, \cite{SokeretalGG2019} argued that we do expect CEJSN events to take place in star-forming galaxies. I here add that the envelope of the RSG star must be massive enough, crudely $M_{\rm 2,env} \ga 10 M_\odot$, to force the NS companion to spiral-in down to final core-NS orbital separation of about $a_{\rm NS} \simeq 1-3 R_\odot$.  This implies that not only the progenitor of the NS was a massive star, but the initially less massive star, the progenitor of the RSG star, should also be a massive star, i.e., with an initial mass of $\ga 8 M_\odot$. However, in the present scenario the RSG envelope cannot be too massive as to force the NS to spiral all the way to the core. I return to my suggestion that some FBOTs are progenitors of binary NSs that later merge in section \ref{subsec:Rate}. 

\subsection{Hydrogen in the fast ejecta}
\label{subsec: Hydrogenejecta}
There are two sources to the hydrogen in the fast ejecta.
(1) One of the new ingredients that I add here is that the NS accretes mass from a circumbinary disk when it orbits the core at a very small radii of $a_{\rm NS} \simeq 1-3 R_\odot$ (Section \ref{subsec:CBD}). This circumbinary disk is the leftover of the RSG envelope, hence it is hydrogen-rich. Its inner and outer radii are $R_{\rm CBD,in} \simeq a_{\rm NC}$ and $R_{\rm CBD,out} \simeq 5-30 R_\odot$, respectively, while its mass is crudely $M_{\rm CBD}~\simeq~0.1-1~M_\odot$.
(2) The jets that the NS launches at the explosion as it accretes mass from the circumbinary disk (section \ref{subsec:CBD}; it might accrete some mass from the core of the RSG) sweep hydrogen-rich CSM that was ejected during the CEE of the NS inside the RSG envelope. 

\subsection{High velocities of $v_{\rm ej} > 0.1c$}
\label{subsec: High velocities}
The outflow velocities in the FBOTs AT 2018cow, ZTF18abvkwla, and CRTSCSS161010 J045834-081803 (CSS16 hereafter) are $\simeq 0.1 c$ (e.g., \citealt{Marguttietal2019}), $\simeq 0.3c$ (e.g., \citealt{Hoetal2020}), and  $\simeq 0.5 c$ \citep{Coppejansetal2020}, respectively.  
I attribute these high velocities (relative to CCSNe) to the clean polar directions that allow the jets to expand while interacting with a relatively low envelope and CSM mass, at least during part of the event. 

\cite{Perleyetal2019} find that in AT2018cow the high-velocity absorption lines disappear after about two weeks. I attribute this to that the fast material has expanded to large distances and the dominant absorbing gas is the slower gas at large angle to the symmetry axis. Namely, a gas that comes mainly from the slower walls of the polar lobes (Fig. \ref{fig:FBOTsInteractionfig4}). This suggestion requires further study by hydrodynamical simulations.  

\subsection{A small mass of ejecta at $v_{\rm ej} > 0.1c$}
\label{subsec: SmallMass}
\cite{Coppejansetal2020}, for example, estimate the fast ejecta mass in the FBOT CSS16 to be $\simeq 0.01-0.1 M_\odot$ and the kinetic energy $\simeq 10^{51}-10^{52} \erg$.
The circumbinary disk at the end of the CEE inside the envelope is expected to contain a small fraction of the original envelope mass, and therefore the mass that the jets carry is small, as equation (\ref{eq:JetsMass}) gives. The jets expand almost freely, but not totally so as the lobes cannot be completely empty. There is a gas inside the lobes before explosion, with lower densities and higher temperatures than those of the walls of the lobes (the optically-thick shell).  As such, the jets entrain gas and slows down. For example, in AT2018cow I expect the jets to interact with gas in the lobes that is a few times more massive than the jets, so the jets slow down from $v_j \simeq 10^5 \km \s^{-1}$ at launching to few times slower, $\simeq 0.1c$. 
As well, parts of the jets, in particular at large angles to the symmetry axis, might be chocked by the wider walls of the lobes (the shell) at low latitudes. Overall, the mass in the fast ejecta might crudely be 
\begin{eqnarray}
& M(>0.1 c) \approx 0.3 M_{\rm 2j} -  3 M_{\rm 2j} \nonumber
\\ &   \approx  (0.003 - 0.03 )\times 
\left( \frac{\eta_{2j}}{0.05} \right)
\left( \frac{\eta_{\rm CBD}}{0.02} \right)
\left( \frac{M_{\rm env}}{10 M_\odot} \right) M_\odot,  
\label{eq:Mfast}    
\end{eqnarray}
{ where in the second equality I inserted equation (\ref{eq:JetsMass}). This is compatible with the fast ejecta in the FBOT CSS16 \citep{Coppejansetal2020}, $\simeq 0.01-0.1 M_\odot$, as the value of $\eta_{2j} \eta_{\rm CBD}$ might be larger by up to a factor of few than the scaling of equations (\ref{eq:JetsMass}) and (\ref{eq:Mfast}).
}

\subsection{Total event energy of $\approx 10^{50}-10^{52} \erg$}
\label{subsec:TotalEnergy}
I attribute this energy of $< 10^{52} \erg$ to the accretion from a low mass circumbinary disk, as equation (\ref{eq:JetsEnergy}) shows. I note that in accretion disks of collapsars, i.e., the collapse of the core in a CCSN that forms a BH with an accretion disk around it, nuclear burning with helium might take place (e.g., \citealt{Zenatietal2020}). Here the accretion disk is hydrogen-rich and of lower densities. Future numerical simulations should examine whether nuclear reactions take place in the accretion disk. 

\subsection{A fast rise (a day to few days)} 
\label{subsec:FastRise}

The rise time of AT2018cow, as an example, was less than three days, and at a time of $\Delta t < 1.3 \days$ its magnitude raised by 4.2 mag (\citealt{Prenticeeta2018, Perleyetal2019}). 
I estimate this typical timescale in equation (\ref{eq:tdiff}) as the photon diffusion time. With that comes the observations that AT2018cow-like  transients are the most luminous and fast type of FBOTs (e.g., \citealt{Hoetal2021a}). This requires an efficient channelling of kinetic energy to thermal energy and then radiation. The low densities at large distances account for that (equation \ref{eq:tdiff}).

The shock breakout through the walls of the lobs at large distances accounts for the properties of AT2018cow as \cite{Perleyetal2019} deduced. 
They concluded that at shock breakout the photosphere should be unbound and results from a pre-explosion dense wind or shell ejection. \cite{Perleyetal2019} concluded also that the CSM shell should be localised in extent. The fronts of the lobes (Figs. \ref{fig:FBOTsCSM} - \ref{fig:FBOTsInteractionfig4}) have these properties. 

\subsection{Decreasing photosphere in AT2018cow} 
\label{subsec:Decreasing}
In AT2018cow the photosphere decreases during the first several weeks of observations \citep{Perleyetal2019}. In section \ref{subsec:Interaction} I attributed this structure to the pre-collapse lobes and the nature of the interaction with the jets that removes the fronts of the lobes such that the area of photosphere decreases (schematically in the transition from Fig. \ref{fig:FBOTsInteractionfig3} to Fig. \ref{fig:FBOTsInteractionfig4}).
The complicated structure of the photosphere might account for the increase in photospheric temperature after several weeks in AT2018cow \citep{Perleyetal2019}. Future hydrodynamical numerical simulations will determine the exact behavior. The large parameter space of CSM shapes and densities and the jets' properties will require an intensive study. 

\subsection{Rapidly variable X-ray source} 
\label{subsec:RapidX-ray}
The X-ray emission of AT2018cow-like FBOTs varies from timescales of days (e.g., \citealt{Yaoetal2022} for AT2020mrf) down to a fraction of a second (e.g., \citealt{Pashametal2021} for AT2018cow).
Some X-ray might come from the jets as they pass through shocks in optically-thin regions for X-ray. This might account for hours to days timescales of variability. This is a subject of future hydrodynamical simulations. 

The rapid variability can come from the accretion disk. 
The X-ray emission of FBOTs requires a central compact source (engine), e.g., AT2020xnd \citep{Hoetal2022}. The rapid X-ray variability even weeks after explosion points to a central energy course (engine; e.g., \citealt{Pashametal2021}). The central engine most likely involves jets as in the polar CEJSN scenario \citep{SokeretalGG2019} and in the scenario where the inner parts of a star collapse to form a BH that launches jets (e.g., \citealt{Perleyetal2021, Gottliebetal2022}).  
This is also the case with the polar CEJSN impostor scenario that I propose here where a circumbinary disk feeds the accretion disk around the NS (or BH in some cases) that launches the jets. 
Another central engine is a magnetar (e.g., \citealt{Mohanetal2020}).
However, the formation of an energetic magnetar most likely is accompanied by more energetic jets (e.g., \citealt{SokerGilkis2017}). Note that the tidal disruption scenario (e.g., \citealt{Kuinetal2019, Perleyetal2019}) seems unable to explain FBOTs { because FBOTs come from massive star and have dense CSM } (e.g., \citealt{Huangetal2019, Yaoetal2022}). 

\subsection{Possible dense CSM at $r \simeq 10^{17} \cm$} 
\label{subsec:DenseCSM}
AT2018cow (e.g., \citealt{NayanaChandra2021}) and AT2020xnd (e.g., \citealt{Brightetal2021}) have a dense CSM that extends up to $r \approx 10^{17} \cm$. \cite{Coppejansetal2020} conclude that the ejecta of the FBOT CSS16 interacts with a dense wind at $r \approx 10^{17} \cm$. From the deceleration of the outflow they conclude that the CSM mass that the outflow (blast wave) sweeps is comparable or larger than the mass of the fast material (ejecta). In the polar CEJSN impostor scenario the extended CSM is the CSM halo (Fig. \ref{fig:FBOTsCSM}) that I discussed in section \ref{subsec:CSM}, which extends to $R_{\rm halo} \approx 10^{17} \cm$ { and which, although contains a mass larger than a regular RSG wind, i.e., corresponding to a mass loss rate of $\dot M_{\rm halo} \approx 10^{-3}-10^{-4} M_\odot \yr^{-1}$, is } optically thin. { The close CSM, at $r < R_{\rm L} \approx 10^{15} \cm$, is much more massive $M_{\rm CSM} \ga 10 M_\odot$, and is concentrated in the equatorial plane }. 

\subsection{The rate of AT2018cow-like FBOTs}
\label{subsec:Rate}

\cite{Hoetal2022} estimate the rate of AT2018cow-like events to be $\simeq 0.01-0.1\%$ of the rate of CCSNe (see also \citealt{Hoetal2021a}). In \cite{SokeretalGG2019} we estimate the rate of all polar-CEJSN and impostor events to be $0.2-0.5\%$ of all CCSNe. The polar CEJSN impostors are a fraction of these, and so the expected rate is compatible with the estimate of \cite{Hoetal2022}. The outcome of the polar CEJSN impostor scenario after the core explodes is a binary system of two NSs that might merge at a later time. The NS-NS merger rate is $\approx 1 \%$ of all CCSNe (e.g., see discussion by \citealt{Mapellietal2018}). If these rates hold, then AT2018cow-like events account for $\approx 10 \%$ of the progenitors of NS-NS mergers. However, the rate of all types of FBOTs is $\approx 1 \%$ of all CCSNe (e.g., {Coppejansetal2020}). As well, some CEJSN impostor channels that are similar, but not identical, to those I study here might account for other fast transients, e.g., AT2018lqh that \cite{Ofeketal2021} observed. \cite{Tsunaetal2021} propose a scenario for AT2018lqh where a rotating blue supergiant collapses to form a BH of $\simeq 30 M_\odot$ and blows a disk wind of $\simeq 0.8 M_\odot$. I instead propose that AT2018lqh is a type of CEJSN impostor similar to the polar CEJSN impostors that explain AT2018cow-like transients. 
\cite{Ofeketal2021} estimate the rate of such fast transients to be of the order of magnitude of the the rate of NS-NS merger. 

I raise therefore the possibility that FBOTs and similar fast transients are CEJSN impostors which compose a large fraction of the progenitors of NS-NS merger binaries. 

\subsection{Optical and radio emission }
\label{subsec:OpticalRAdio}

I will not study here the observed radio properties (e.g., \citealt{Huangetal2019, Hoetal2022}) nor the exact optical emission, as these require more detail radiative transfer calculations following hydrodynamical simulations. I referred to the light curve in sections \ref{subsec:Interaction} and \ref{subsec:FastRise}, and to the photosphere in section \ref{subsec:Decreasing}. I also pointed out (section \ref{sec:NewIngredients}) that I expect the emission properties to be similar to some of those in the studies of \cite{Marguttietal2019} and \cite{Gottliebetal2022}. 

The common properties of the polar CEJSN impostor scenario and of the geometrical model of \cite{Marguttietal2019} are the presence of a dense equatorial outflow, which can explain the late outflow velocity of $\approx 1000 - {\rm several} \times 1000 \km \s^{-1}$ in AT2018cow, the possibility that the ejecta shocks this dense equatorial outflow to contribute to the X-ray emission, and the opening view along the polar directions that allows an observer to view the central engine. 
The common properties with the scenario that \cite{Gottliebetal2022} propose are 
the hot cocoons along the two polar directions that explain the early optical emission and the jets along the polar directions that might contribute to the X-ray emission even at late times. 

{To summarize this section I list the properties and processes that I discussed above} in Table \ref{tab:Table1}. I note that \cite{SokeretalGG2019} already discussed some of these properties and processes in the frame of the polar CEJSN scenario, i.e., for which the NS enters the RSG core. 
\begin{table*}
\centering
\begin{tabular}{|p{0.25\textwidth}|p{0.45\textwidth}|p{0.23\textwidth}|}
\hline
 AT2018cow-like property  & The polar CEJSN impostor scenario  & Typical scenario values \\ 
\hline
\hline
FBOTs tend to occur in star-forming galaxies. (\S \ref{subsec:Star-forming galaxies}) & The secondary star envelope must be sufficiently massive to force the NS/BH to spiral deep down to small $a_{\rm NC}$. 
     & $a_{\rm NC} \simeq 1-3 R_\odot;\qquad \qquad$  $M_{\rm 2,env} \ga 10 M_\odot$. \\ 
\hline
 Hydrogen in the fast outflow. (\S \ref{subsec: Hydrogenejecta}) & (1) The fast ejecta (jets) sweeps CSM mass. (2) A post-CEE circumbinary disk from the inner envelope of the RSG secondary star feeds the jets.  
   & $R_{\rm CBD,in} \simeq a_{\rm NC}; \qquad $  $R_{\rm CBD,out} \simeq 5-30 R_\odot; \qquad$ $M_{\rm CBD}~\simeq~0.1-1~M_\odot$.\\
\hline
High velocities of $v_{\rm ej} > 0.1c$. (\S \ref{subsec: High velocities}) &  The polar directions of the envelope and the CSM are almost empty, allowing the jets an almost free expansion. & To be determined by hydrodynamical simulations. \\
\hline
A small mass of ejecta expands at $>0.1 c$. (\S \ref{subsec: SmallMass}) & The mass in the jets $M_{\rm 2j}$ is given by equation (\ref{eq:JetsMass}). However, the fast jets ($v_{\rm j} \gg 0.1$) can entrain CSM gas. On the other hand, some of the jet mass might be chocked. &  $M(>0.1 c) \approx 0.3 M_{\rm 2j} -  3 M_{\rm 2j}$.\\ 
\hline
Total event energy of $\simeq 10^{50} - 10^{52} \erg$. (\S \ref{subsec:TotalEnergy}) & This is the energy that the jets carry. Most of the accretion energy is carried by neutrinos.   &  Equation (\ref{eq:JetsEnergy}).  \\   
\hline
A fast rise (a day to few days). (\S \ref{subsec:FastRise}) & The shock breaks out from the CSM at $\simeq R_{\rm L} \gg R_{\rm RSG}$ and the interaction region contains a small amount of mass $M_{\rm I}$. Both lead to a short photon diffusion time $t_{\rm diff}$. & $t_{\rm diff} \approx {\rm days}$ (equation \ref{eq:tdiff}). \\
\hline
Decreasing photosphere in AT2018cow. (\S \ref{subsec:Decreasing}) &  The jets-lobes interaction might have a decreasing photosphere area. Fig. \ref{fig:FBOTsInteractionfig3} to Fig. \ref{fig:FBOTsInteractionfig4} qualitatively shows this evolution. & To be determined by hydrodynamical simulations.      
                   \\
\hline
Rapidly variable X-ray source. (\S \ref{subsec:RapidX-ray}) &  Hours to days variability might result from shocks of the jets. Variability with timescales of seconds and below comes from the accretion disk. & To be determined by hydrodynamical simulations.      
\\
\hline
Dense CSM at $\approx 10^{15}-10^{17} \cm$. (\S \ref{subsec:DenseCSM}) & Pre-CEE RSG wind forms the extended optically-thin halo (Fig. \ref{fig:FBOTsCSM}). Jets near the onset of the CEE forms the lobes. The CEE ejects most of the RSG envelope near the equatorial plane starting $t_{\rm CSM} \approx 1-10 \yr$  before explosion. 
     &  $M_{\rm CSM} \ga 10 M_\odot; \qquad \qquad $ $R_{\rm L} \approx 10^{15} \cm; \qquad \qquad$ $R_{\rm halo} \approx 10^{16} - 10^{17} \cm. \qquad$ Section \ref{subsec:CSM}.  \\
\hline
Rate of all FBOTs $\approx 1 \%$ of the CCSN rate, and those of AT~2018cow-like events $<0.1\%$ of the CCSN rate. (\S \ref{subsec:Rate}) & Rate of all polar-CEJSNe { together with } polar-CEJSNe impostors is $\simeq 0.2-0.5\%$ of all CCSNe \citep{SokeretalGG2019}. Polar CEJSN impostors comprise a fraction of that and therefore might account for all AT~2018cow-like events $+$ some other FBOTs. & To be refined by population synthesis studies.  
                   \\
\hline
Optical and radio properties. (\S \ref{subsec:OpticalRAdio}) & Early emission from the hot cocoon (Fig. \ref{fig:FBOTsInteractionfig3}) similar in some aspects to the hot-cocoon in the scenario of \cite{Gottliebetal2022}. Late emission from the dense equatorial region similar in some aspects to that in the geometrical model of \cite{Marguttietal2019}. & Quantitative study requires hydrodynamical simulations + radiative transfer calculations.       
                   \\                   
                   \hline
\end{tabular}
\caption{The explanation of the polar-CEJSN impostor scenario to the different properties of AT2018cow-like FBOTs. Note that in some cases I propose explanations that require detailed simulations and calculations.  
}
\label{tab:Table1}
\end{table*}

Further hydrodynamical simulations together with radiative transfer calculations are required to determine the optical and radio properties of the polar CEJSN impostor scenario.   
  
\section{Summary}
\label{sec:Summary}

{ The puzzling properties of FBOTs, like fast rise and decline, X-ray variability, and fast outflows, have lead different researchers to propose several different theoretical scenarios to account for FBOTs, in particular AT2018cow-like FBOTs (see references in section \ref{sec:intro}). In the present study I added another scenario, the polar CEJSN impostor scenario. I described the main evolutionary phases of this scenario in section \ref{sec:PolarImpostor}.  }

{ The main new ingredients of the polar CEJSN impostor scenario are a pre-explosion bipolar CSM (section \ref{subsec:CSM}, Fig. \ref{fig:FBOTsCSM}), the post-CEE accretion from a hydrogen-rich circumbinary disk that feeds the accretion disk around the NS that launches the jets (section \ref{subsec:CBD}), and the interaction of these jets with the lobes (section \ref{subsec:Interaction}; Figs. \ref{fig:FBOTsInteractionfig2}-\ref{fig:FBOTsInteractionfig4}). The main energy source of FBOTs in the scenario I propose is the gravitational energy of the accretion process onto a pre-existing NS that takes place immediately after the termination of the CEE. The NS accretes mass from an accretion disk that launches jets. The post-CEE circumbinary disk feeds the accretion disk for a time period of weeks (section \ref{sec:PolarImpostor}; Fig. \ref{fig:FBOTsCSM}). The collision of the jets with the CSM give rise to thermal energy and then radiation (Figs. \ref{fig:FBOTsInteractionfig2}-\ref{fig:FBOTsInteractionfig4}). }

{ In section \ref{sec:ObservatinalProperties} I listed different properties of AT2018cow-like FBOTs and discussed the way by which the polar CEJSN impostor scenario might account for these properties (Table \ref{tab:Table1}). }

{ At the end of the FBOT there is a bare system of a NS and the core of the RSG. The core later explodes, leading to the formation of a binary NS system that might be bound. Such a binary might much later experience NS-NS merger. 
In section \ref{subsec:Rate} I crudely estimated, based on the proposed scenario, that AT2018cow-like events are progenitors of $\approx 10 \%$ of the NS-NS mergers. From that I raised the possibility that FBOTs and similar fast transients are CEJSN impostors that are progenitors of a large fraction of NS-NS merger binaries. }

\section*{Acknowledgments}

I thank Ore Gottlieb and Aldana Grichener for useful comments, 
{and an anonymous referee for detailed comments that substantially improved the manuscript. } 
This research was supported by the Amnon Pazy Research Foundation and the Asher Space Research Fund at the Technion.



\label{lastpage}

\begin{thebibliography}{}\addcontentsline{toc}{section}{References}

\bibitem[\protect\citeauthoryear{Akashi \& Soker}{2008}]{AkashiSoker2008} Akashi M., Soker N., 2008, MNRAS, 391, 1063. doi:10.1111/j.1365-2966.2008.13935.x

\bibitem[Arcavi et al.(2017)]{Arcavietal2017} Arcavi, I., Howell, D.~A., Kasen, D., et al.\ 2017, \nat, 551, 210
 
\bibitem[Barkov \& Komissarov(2011)]{BarkovKomissarov2011} Barkov, M.~V., \& Komissarov, S.~S.\ 2011, \mnras, 415, 944

\bibitem[\protect\citeauthoryear{Bietenholz et al.}{2020}]{Bietenholzet2020B} Bietenholz M.~F., Margutti R., Coppejans D., Alexander K.~D., Argo M., Bartel N., Eftekhari T., et al., 2020, MNRAS, 491, 4735. doi:10.1093/mnras/stz3249

\bibitem[\protect\citeauthoryear{Bright et al.}{2021}]{Brightetal2021} Bright J.~S., Margutti R., Matthews D., Brethauer D., Coppejans D., Wieringa M.~H., Metzger B.~D., et al., 2021, arXiv, arXiv:2110.05514

\bibitem[\protect\citeauthoryear{Chen \& Shen}{2022}]{ChenShen2022} { Chen C., Shen R.-F., 2022, arXiv:2201.12534 }

\bibitem[Chevalier(2012)]{Chevalier2012} Chevalier, R.~A.\ 2012, \apjl, 752, L2 

\bibitem[\protect\citeauthoryear{Cohen et al.}{1975}]{Cohenetal1975} Cohen M., Anderson C.~M., Cowley A., Coyne G.~V., Fawley W., Gull T.~R., Harlan E.~A., et al., 1975, ApJ, 196, 179. doi:10.1086/153403

\bibitem[\protect\citeauthoryear{Coppejans et al.}{2020}]{Coppejansetal2020} Coppejans D.~L., Margutti R., Terreran G., Nayana A.~J., Coughlin E.~R., Laskar T., Alexander K.~D., et al., 2020, ApJL, 895, L23. doi:10.3847/2041-8213/ab8cc7

\bibitem[\protect\citeauthoryear{Corradi et al.}{2003}]{Corradietal2003} Corradi R.~L.~M., Sch{\"o}nberner D., Steffen M., Perinotto M., 2003, MNRAS, 340, 417. doi:10.1046/j.1365-8711.2003.06294.x
  
\bibitem[\protect\citeauthoryear{Dong et al.}{2021}]{Dongetal2021} Dong D.~Z., Hallinan G., Nakar E., Ho A.~Y.~Q., Hughes A.~K., Hotokezaka K., Myers S.~T., et al., 2021, Sci, 373, 1125. doi:10.1126/science.abg6037

\bibitem[\protect\citeauthoryear{Fox \& Smith}{2019}]{FoxSmith2019} Fox O.~D., Smith N., 2019, MNRAS, 488, 3772. doi:10.1093/mnras/stz1925
    
\bibitem[Fryer \& Woosley(1998)]{FryerWoosley1998} Fryer, C.~L., \& Woosley, S.~E.\ 1998, \apjl, 502, L9

\bibitem[\protect\citeauthoryear{Garc{\'\i}a-D{\'\i}az et al.}{2018}]{GarciaDiazetal2018} Garc{\'\i}a-D{\'\i}az M.~T., Steffen W., Henney W.~J., L{\'o}pez J.~A., Garc{\'\i}a-L{\'o}pez F., Gonz{\'a}lez-Buitrago D., {\'A}viles A., 2018, MNRAS, 479, 3909. doi:10.1093/mnras/sty1590

\bibitem[Gilkis et al.(2019)]{Gilkisetal2019} Gilkis, A., Soker, N., \& Kashi, A.\ 2019, \mnras, 482, 4233. doi:10.1093/mnras/sty3008

\bibitem[\protect\citeauthoryear{Glanz \& Perets}{2021}]{GlanzPerets2021} Glanz H., Perets H.~B., 2021, MNRAS, 500, 1921. doi:10.1093/mnras/staa3242
 
\bibitem[\protect\citeauthoryear{Gottlieb, Tchekhovskoy, \& Margutti}{2022}]{Gottliebetal2022} Gottlieb O., Tchekhovskoy A., Margutti R., 2022, arXiv:2201.04636


\bibitem[\protect\citeauthoryear{Grichener, Cohen, \& Soker}{2021}]{Gricheneretal2021} Grichener A., Cohen C., Soker N., 2021, ApJ, 922, 61. doi:10.3847/1538-4357/ac23dd

\bibitem[\protect\citeauthoryear{Grichener, Kobayashi, \& Soker}{2022}]{Gricheneretal2022} Grichener A., Kobayashi C., Soker N., 2022, arXiv, arXiv:2112.08301

\bibitem[Grichener \& Soker(2019a)]{GrichenerSoker2019a} Grichener, A., \& Soker, N.\ 2019a, \apj, 878, 24

\bibitem[Grichener \& Soker(2019b)]{GrichenerSoker2019b} Grichener, A. \& Soker, N.\ 2019b, arXiv:1909.06328

\bibitem[\protect\citeauthoryear{Grichener \& Soker}{2021}]{GrichenerSoker2021} Grichener A., Soker N., 2021, MNRAS, 507, 1651. doi:10.1093/mnras/stab2233

\bibitem[\protect\citeauthoryear{Han et al.}{2020}]{Hanetal2020} { Han Z.-W., Ge H.-W., Chen X.-F., Chen H.-L., 2020, RAA, 20, 161. doi:10.1088/1674-4527/20/10/161 }

\bibitem[\protect\citeauthoryear{Hillel, Schreier, \& Soker}{2021}]{Hilleletal2022} Hillel S., Schreier R., Soker N., 2022, arXiv, arXiv:2112.01459

\bibitem[\protect\citeauthoryear{Ho et al.}{2022}]{Hoetal2022} Ho A.~Y.~Q., Margalit B., Bremer M., Perley D.~A., Yao Y., Dobie D., Kaplan D.~L., et al., 2022, arXiv, arXiv:2110.05490

\bibitem[\protect\citeauthoryear{Ho et al.}{2021}]{Hoetal2021a} Ho A.~Y.~Q., Perley D.~A., Gal-Yam A., Lunnan R., Sollerman J., Schulze S., Das K.~K., et al., 2021, arXiv, arXiv:2105.08811

\bibitem[\protect\citeauthoryear{Ho et al.}{2020}]{Hoetal2020} Ho A.~Y.~Q., Perley D.~A., Kulkarni S.~R., Dong D.~Z.~J., De K., Chandra P., Andreoni I., et al., 2020, ApJ, 895, 49. doi:10.3847/1538-4357/ab8bcf

\bibitem[\protect\citeauthoryear{Huang et al.}{2019}]{Huangetal2019} Huang K., Shimoda J., Urata Y., Toma K., Yamaoka K., Asada K., Nagai H., et al., 2019, ApJL, 878, L25. doi:10.3847/2041-8213/ab23fd

\bibitem[\protect\citeauthoryear{Kashi \& Soker}{2011}]{KashiSoker2011} Kashi A., Soker N., 2011, MNRAS, 417, 1466. doi:10.1111/j.1365-2966.2011.19361.x

\bibitem[\protect\citeauthoryear{Kremer et al.}{2021}]{Kremeretal2021} Kremer K., Lu W., Piro A.~L., Chatterjee S., Rasio F.~A., Ye C.~S., 2021, ApJ, 911, 104. doi:10.3847/1538-4357/abeb14

\bibitem[\protect\citeauthoryear{Kuin et al.}{2019}]{Kuinetal2019} Kuin N.~P.~M., Wu K., Oates S., Lien A., Emery S., Kennea J.~A., de Pasquale M., et al., 2019, MNRAS, 487, 2505. doi:10.1093/mnras/stz053

\bibitem[\protect\citeauthoryear{Lau et al.}{2022}]{Lauetal2022} { Lau M.~Y.~M., Hirai R., Gonz{\'a}lez-Bol{\'\i}var M., Price D.~J., De Marco O., Mandel I., 2022, MNRAS.tmp. doi:10.1093/mnras/stac049 }

\bibitem[\protect\citeauthoryear{Leung et al.}{2020}]{Leungetal2020} Leung S.-C., Blinnikov S., Nomoto K., Baklanov P., Sorokina E., Tolstov A., 2020, ApJ, 903, 66. doi:10.3847/1538-4357/abba33

\bibitem[\protect\citeauthoryear{Liu et al.}{2018}]{Liuetal2018} Liu L.-D., Zhang B., Wang L.-J., Dai Z.-G., 2018, ApJL, 868, L24. doi:10.3847/2041-8213/aaeff6

\bibitem[L{\'o}pez-C{\'a}mara et al.(2019)]{LopezCamaraetal2019} L{\'o}pez-C{\'a}mara, D., De Colle, F., \& Moreno M{\'e}ndez, E.\ 2019, \mnras, 482, 3646

\bibitem[L{\'o}pez-C{\'a}mara et al.(2020)]{LopezCamaraetal2020MN} L{\'o}pez-C{\'a}mara, D., Moreno M{\'e}ndez, E., \& De Colle, F.\ 2020, \mnras, 497, 2057

\bibitem[\protect\citeauthoryear{Lyman et al.}{2020}]{Lymanetal2020} Lyman J.~D., Galbany L., S{\'a}nchez S.~F., Anderson J.~P., Kuncarayakti H., Prieto J.~L., 2020, MNRAS, 495, 992. doi:10.1093/mnras/staa1243

\bibitem[\protect\citeauthoryear{Lyutikov \& Toonen}{2019}]{LyutikovToonen2019} Lyutikov M., Toonen S., 2019, MNRAS, 487, 5618. doi:10.1093/mnras/stz1640

\bibitem[\protect\citeauthoryear{Mapelli et al.}{2018}]{Mapellietal2018} Mapelli M., Giacobbo N., Toffano M., Ripamonti E., Bressan A., Spera M., Branchesi M., 2018, MNRAS, 481, 5324. doi:10.1093/mnras/sty2663

\bibitem[\protect\citeauthoryear{Margutti et al.}{2019}]{Marguttietal2019} Margutti R., Metzger B.~D., Chornock R., Vurm I., Roth N., Grefenstette B.~W., Savchenko V., et al., 2019, ApJ, 872, 18. doi:10.3847/1538-4357/aafa01

\bibitem[\protect\citeauthoryear{Mohan, An, \& Yang}{2020}]{Mohanetal2020} Mohan P., An T., Yang J., 2020, ApJL, 888, L24. doi:10.3847/2041-8213/ab64d1

\bibitem[\protect\citeauthoryear{Nayana \& Chandra}{2021}]{NayanaChandra2021} Nayana A.~J., Chandra P., 2021, ApJL, 912, L9. doi:10.3847/2041-8213/abed55

\bibitem[\protect\citeauthoryear{Ofek et al.}{2021}]{Ofeketal2021} Ofek E.~O., Adams S.~M., Waxman E., Sharon A., Kushnir D., Horesh A., Ho A., et al., 2021, ApJ, 922, 247. doi:10.3847/1538-4357/ac24fc

\bibitem[Papish et al.(2015)]{Papishetal2015} Papish, O., Soker, N., \& Bukay, I.\ 2015, \mnras, 449, 288

\bibitem[\protect\citeauthoryear{Pasham et al.}{2021}]{Pashametal2021} Pasham D.~R., Ho W.~C.~G., Alston W., Remillard R., Ng M., Gendreau K., Metzger B.~D., et al., 2021, arXiv, arXiv:2112.04531

\bibitem[\protect\citeauthoryear{Perley et al.}{2021}]{Perleyetal2021} Perley D.~A., Ho A.~Y.~Q., Yao Y., Fremling C., Anderson J.~P., Schulze S., Kumar H., et al., 2021, MNRAS, 508, 5138. doi:10.1093/mnras/stab2785

\bibitem[\protect\citeauthoryear{Perley et al.}{2019}]{Perleyetal2019} Perley D.~A., Mazzali P.~A., Yan L., Cenko S.~B., Gezari S., Taggart K., Blagorodnova N., et al., 2019, MNRAS, 484, 1031. doi:10.1093/mnras/sty3420

\bibitem[\protect\citeauthoryear{Piro \& Lu}{2020}]{PiroLu2020} Piro A.~L., Lu W., 2020, ApJ, 894, 2. doi:10.3847/1538-4357/ab83f6

\bibitem[\protect\citeauthoryear{Prentice et al.}{2018}]{Prenticeeta2018} Prentice S.~J., Maguire K., Smartt S.~J., Magee M.~R., Schady P., Sim S., Chen T.-W., et al., 2018, ApJL, 865, L3. doi:10.3847/2041-8213/aadd90

\bibitem[\protect\citeauthoryear{Quataert, Lecoanet, \& Coughlin}{2019}]{Quataertetal2019} Quataert E., Lecoanet D., Coughlin E.~R., 2019, MNRAS, 485, L83. doi:10.1093/mnrasl/slz031

\bibitem[\protect\citeauthoryear{Schreier et al.}{2021}]{Schreieretal2021} Schreier R., Hillel S., Shiber S., Soker N., 2021, MNRAS, 508, 2386. doi:10.1093/mnras/stab2687

\bibitem[Schr{\o}der et al.(2020)]{Schroderetal2020} Schr{\o}der, S.~L., MacLeod, M., Loeb, A., et al.\ 2020, \apj, 892, 13

\bibitem[\protect\citeauthoryear{Shiber et al.}{2019}]{Shiberetal2019} Shiber S., Iaconi R., De Marco O., Soker N., 2019, MNRAS, 488, 5615. doi:10.1093/mnras/stz2013


\bibitem[\protect\citeauthoryear{Shiber, Schreier, \& Soker}{2016}]{Shiberetal2016} { Shiber S., Schreier R., Soker N., 2016, RAA, 16, 117. doi:10.1088/1674-4527/16/7/117 }

\bibitem[\protect\citeauthoryear{Smith}{2006}]{Smith2006} Smith N., 2006, ApJ, 644, 1151. doi:10.1086/503766

\bibitem[\protect\citeauthoryear{Soker}{1992}]{Soker1992} Soker N., 1992, ApJ, 389, 628. doi:10.1086/171235

\bibitem[\protect\citeauthoryear{Soker}{2019}]{Soker2019TerminateCEE} { Soker N., 2019, MNRAS, 483, 5020. doi:10.1093/mnras/sty3496 }

\bibitem[\protect\citeauthoryear{Soker}{2021}]{Soker2021} Soker N., 2021, MNRAS, 504, 5967. doi:10.1093/mnras/stab1275

\bibitem[\protect\citeauthoryear{Soker \& Gilkis}{2017}]{SokerGilkis2017} Soker N., Gilkis A., 2017, ApJ, 851, 95. doi:10.3847/1538-4357/aa9c83

\bibitem[Soker \& Gilkis(2018)]{SokerGilkis2018} Soker, N., \& Gilkis, A.\ 2018, \mnras, 475, 1198

\bibitem[\protect\citeauthoryear{Soker, Grichener, \& Gilkis}{2019}]{SokeretalGG2019} Soker N., Grichener A., Gilkis A., 2019, MNRAS, 484, 4972. doi:10.1093/mnras/stz364

\bibitem[\protect\citeauthoryear{Tauris, Langer, \& Podsiadlowski}{2015}]{Taurisetal2015} Tauris T.~M., Langer N., Podsiadlowski P., 2015, MNRAS, 451, 2123. doi:10.1093/mnras/stv990

\bibitem[Th{\"o}ne et al.(2011)]{Thoneetal2011} Th{\"o}ne, C.~C., de Ugarte Postigo, A., Fryer, C.~L., Page, K.~L., Gorosabel, J., Aloy, M.~A., Perley, D.~A., et al., 2011, Natur, 480, 72. doi:10.1038/nature10611

\bibitem[\protect\citeauthoryear{Tsuna, Kashiyama, \& Shigeyama}{2021}]{Tsunaetal2021} Tsuna D., Kashiyama K., Shigeyama T., 2021, ApJL, 922, L34. doi:10.3847/2041-8213/ac3997

\bibitem[\protect\citeauthoryear{Uno \& Maeda}{2020}]{UnoMaeda2020} Uno K., Maeda K., 2020, ApJ, 897, 156. doi:10.3847/1538-4357/ab9632

\bibitem[\protect\citeauthoryear{Vigna-G{\'o}mez et al.}{2018}]{VignaGomezetal2018} Vigna-G{\'o}mez A., Neijssel C.~J., Stevenson S., Barrett J.~W., Belczynski K., Justham S., de Mink S.~E., et al., 2018, MNRAS, 481, 4009. doi:10.1093/mnras/sty2463

\bibitem[\protect\citeauthoryear{Xiang et al.}{2021}]{Xiangetal2021} Xiang D., Wang X., Lin W., Mo J., Lin H., Burke J., Hiramatsu D., et al., 2021, ApJ, 910, 42. doi:10.3847/1538-4357/abdeba

\bibitem[\protect\citeauthoryear{Yang et al.}{2021}]{Yangetal2020} Yang S., Sollerman J., Chen T.-W., Kool E.~C., Lunnan R., Schulze S., Strotjohann N., et al., 2021, A\&A, 646, A22. doi:10.1051/0004-6361/202039440

\bibitem[\protect\citeauthoryear{Yao et al.}{2022}]{Yaoetal2022} Yao Y., Ho A.~Y.~Q., Medvedev P., Nayana A.~J., Perley D.~A., Kulkarni S.~R., Chandra P., et al., 2022, arXiv, arXiv:2112.00751

\bibitem[Yu et al.(2019)]{Yuetal2019} Yu, Y.-W., Chen, A., \& Li, X.-D.\ 2019, \apjl, 877, L21. doi:10.3847/2041-8213/ab1f85

\bibitem[\protect\citeauthoryear{Zenati et al.}{2020}]{Zenatietal2020} Zenati Y., Siegel D.~M., Metzger B.~D., Perets H.~B., 2020, MNRAS, 499, 4097. doi:10.1093/mnras/staa3002

\bibitem[Zhang \& Fryer(2001)]{ZhangFryer2001} Zhang, W., \& Fryer, C.~L.\ 2001, \apj, 550, 357

\bibitem[\protect\citeauthoryear{Zou et al.}{2020}]{Zouetal2020} Zou Y., Frank A., Chen Z., Reichardt T., De Marco O., Blackman E.~G., Nordhaus J., et al., 2020, MNRAS, 497, 2855. doi:10.1093/mnras/staa2145


\end{thebibliography}
\end{document}